\documentclass[prb,twocolumn, times, aps, amssymb,footinbib,showpacs, superscriptaddress]{revtex4-1}
\usepackage{graphicx}				% Use pdf, png, jpg, or eps§ with pdflatex; use eps in DVI mode
\usepackage{amssymb, amsmath, braket, dsfont, hyperref}
\usepackage{color}
\usepackage{soul}

\newcommand{\bex}{b_{\textrm{ex}}}
\newcommand{\fex}{f_{\textrm{ex}}}
\newcommand{\nf}{\psi_{\textrm{NF}}}
\newcommand{\bqh}{b_{\textrm{QH}}}

\usepackage[dvipsnames]{xcolor}
\definecolor{Zcolour}{rgb}{0.992, 0.588, 0.22}
\definecolor{dkgreen}{rgb}{0,0.5,0}
\definecolor{purple}{rgb}{0.5,0,0.5}

\begin{document}
\title{ Evidence for a topological  ``exciton Fermi sea'' in bilayer graphene}
\author{Michael P. Zaletel}
\affiliation{Department of Physics, Princeton University, Princeton, NJ 08540, U.S.A.}
\author{Scott Geraedts}
\affiliation{Department of Physics, Princeton University, Princeton, NJ 08540, U.S.A.}
\author{Zlatko Papi\'c}
\affiliation{School of Physics and Astronomy, University of Leeds, Leeds, LS2 9JT, United Kingdom}
\author{Edward H. Rezayi}
\affiliation{Department of Physics, California State University, Los Angeles, CA 90032, USA}

\begin{abstract}
The quantum Hall physics of bilayer graphene is extremely rich due to the interplay between a layer degree of freedom and delicate fractional states.
Recent experiments show that when an electric field perpendicular to the bilayer causes Landau levels of opposing  layers to cross in energy, a  even-denominator Hall plateau can coexist with a finite density of inter-layer excitons.
We present theoretical and numerical evidence that this observation is due to a new phase of matter - a Fermi sea of topological excitons.
\end{abstract}
\maketitle
\tableofcontents

\vspace{4mm}

\section{Introduction}

In a topological phase of matter, quasiparticles can emerge with quantum numbers and statistics which are a fraction of the electrons'~\cite{Laughlin}.
While the fractionalization of charge has a number of  dramatic experimental consequences, for example the shot-noise signatures of the charge $e^\ast = \frac{e}{q}$ quasiparticles of the fractional quantum Hall (FQH) effect,\cite{de1997direct, Saminadayar} detecting the fractionalization of statistics is more subtle.
A case of particular interest is ``charge-statistics'' separation: the electron ``$c$'' may fractionalize into a charge $-e$ boson ``$b$'' and a neutral fermion ``$\nf$'',  $c = b \, \nf$, as has been proposed to occur in systems ranging from spin-liquid phases of Mott insulators\cite{Kivelson87, Baskaran87} to mixed-valence insulators \cite{Chowdhury2017} and certain FQH effects.\cite{MooreRead}
Charge-statistics fractionalization is an enticing possibility from an experimental standpoint: if the neutral fermions can be doped to finite density, they may form a ``neutral Fermi surface'' with dramatic  signatures such as quantum  and Friedel oscillations in an electrical insulator. \cite{LeeNagaosa92, Motrunich2006, MrossSenthil2011, barkeshli2014coherent}

One long-standing candidate for charge-statistics fractionalization is the even-denominator FQH effect observed in the $\nu=5/2$-plateau of GaAs,\cite{Willett87} or more recently, the $\nu = \pm \frac{1}{2}$-plateau of Bernal-stacked bilayer graphene (BLG).\cite{Morpurgo2013, Zibrov, li2017even}
Theoretical work suggests these states are a type of ``Pfaffian'' phase featuring non-Abelian anyons.   \cite{MooreRead, Greiter1992, Morf1998, HaldaneRezayi2000, ApalkovChakraborty, PapicAbanin2014, Zibrov}
In the composite fermion (CF) picture of these phases, each electron binds with two magnetic flux quanta to form a composite fermion which  experiences zero net magnetic field. \cite{Jain1989, HLR}
Depending on the interactions, the CFs may condense into a chiral superconductor, opening up a quantized Hall gap with $\sigma_{xy} = \frac{1}{2} \frac{e^2}{h}$. Charge-statistics fractionalization is central to the Pfaffian phase: the boson $b$ is realized as a quadruple-vortex in the CF condensate, which carries charge $-e$ due to the Hall conductance, while the neutral fermion $\nf$ arises as the Bogoliubov-de Gennes (BdG) excitation of a broken CF Cooper pair.\cite{MooreRead, Read2000}
A recent experiment has found intriguing evidence for the existence of the $\nf$  through its contribution to the  quantized thermal-Hall effect of the edge.~\cite{Banerjee2017}

	An interesting question  arises: can the $\nf$ be induced to finite density in order to provide experimental evidence for the putative charge-statistics fractionalization of the even-denominator plateau? Since the $\nf$ carry neither spin nor charge, there is no obvious way to do so.
It was recently argued \cite{Barkeshli2016} - and perhaps even shown experimentally\cite{Zibrov} - that the interplay of an even-denominator quantum Hall effect and valley degeneracy in BLG provides an exciting platform for this purpose.
BLG is formed from two atomically-close layers of graphene, and features quadratic band touchings (valleys) at momenta  $K_-, K_+$~\cite{McCannBLGReview}. In a strong magnetic field, electrons in valley $+$ and $-$ are localized onto the top and bottom layer of the BLG respectively, and tunneling between the two is suppressed. When a perpendicular electric field localizes the electrons onto one layer, a $\nu = \pm \frac{1}{2}$ FQH state is observed.\cite{Morpurgo2013, Zibrov}
As the electric field is reduced, it becomes favorable for charge to distribute onto the opposing layer. Because the equivalence between layer and valley prevents direct hybridization between them, incomplete layer polarization should induce a finite density of long-lived interlayer excitons.
Remarkably, based on capacitive measurements sensitive to the layer polarization, Zibrov \emph{et al.}  \cite{Zibrov} found evidence for the existence of an intermediate phase in which the even-denominator QH gap coexists with partial layer polarization.
The coexistence of an even-denominator gap with a finite density of interlayer excitons has yet to be understood.

In a conventional system, the charge (neutral) and statistics (bosonic) of an exciton is the sum of its parts, and hence at a finite density the excitons could form a bosonic condensate, as has been observed experimentally in integer QH bilayers~\cite{Spielman, tutuc2004counterflow} (see also the recent review~\onlinecite{EisensteinReview}).
However, Ref.~\onlinecite{Barkeshli2016}  pointed out that in systems with charge-statistics fractionalization \emph{fermionic} excitons can form; these can be understood as a composite of the conventional exciton and the neutral fermion $\nf$. 
If the lowest energy excitons are fermions, then at finite density they could instead form a neutral Fermi surface (FS), resulting in an ``exciton metal'' in which a $\sigma_{xy} = \frac{1}{2} \frac{e^2}{h}$ charge gap coexists with finite layer polarizability.
Surprisingly, Ref.~\onlinecite{Barkeshli2016} found numerical evidence that the lowest energy exciton in BLG was indeed a fermion, raising the possibility that the intermediate phase observed experimentally might be an exciton metal.
Since the excitons carry layer polarization, the exciton metal would feature striking transport  phenomena, such as a metallic counterflow resistance in an electrical insulator, which would provide a new type of evidence for charge-statistics fractionalization.

	The possibility of an exciton metal in BLG seems extremely exotic, and thus far there has not been a microscopic picture of why fermionic excitons should form, or whether their interactions would be favorable to the formation of a Fermi surface.
In this work we use large-scale exact diagonalization (ED) and density matrix renormalization group (DMRG) calculations to model the BLG system, and find compelling evidence for an exciton metal, in support of  Ref.~\onlinecite{Barkeshli2016}.
Furthermore, we show that this seemingly exotic object actually forms for simple electrostatic reasons, due to the peculiar shape of Landau level (LL) wavefunctions in BLG. Together, our results imply this exotic fractionalized metal is a realistic candidate for the intermediate phase observed in experiment.

We begin by reviewing the BLG setup which was explored experimentally in Ref.~\onlinecite{Zibrov}, as well as the theoretical proposal for an exciton metal laid out in Ref.~\onlinecite{Barkeshli2016} (Sec.~\ref{sec:review}).
We then present a microscopic picture of the fermionic exciton which explains its  stability (Sec.~\ref{sec:picture}).
To understand the properties of these excitons at finite density, we  use  exact diagonalization to study the  two-layer QH system relevant to the BLG experiments (Sec.~\ref{sec:ed}).
Starting from the layer-polarized Pfaffian phase, we induce a small number of excitons by transferring charge onto the opposing layer.  The change in the angular momentum of the ground state with increasing exciton number shows a ``shell filling'' \cite{RezayiReadCFL} effect which indicates the formation of an exciton Fermi surface.
We then attack the problem using DMRG on infinite cylinders (Sec.~\ref{sec:dmrg}).
The behavior of the ground state energy and correlation functions as a function of the layer polarization supports the existence of an intermediate phase which is a charge insulator with gapless excitons.
In contrast to analogous numerical experiments at integer filling (which does not have charge-statistics separation), the correlation functions show no indication of the off-diagonal long range order that would characterize a bosonic exciton condensate.
We conclude with some questions for future work.

\section{Experimental scenario, model, and theoretical proposal \label{sec:review}}

\subsection{Valley crossings in BLG}
The basic ingredient for the exciton metal is a level crossing between a lowest ($N=0$) and first excited ($N=1$) Landau level in the absence of tunneling between them.
The crossing arises in BLG as follows. 
In a magnetic field, the single particle states of BLG collapse into flat Landau levels (LLs)  labeled by their valley ($\xi = \pm$), spin ($\sigma = \uparrow / \downarrow$) and LL index ($N$)~\cite{McCannBLGReview}. The $N=0$ and $N=1$ LLs have \emph{approximately} zero energy, while the higher $|N| > 1$ LLs are split away by a large cyclotron gap, leading to $2 \times 2 \times 2 = 8$ LLs in the low-energy manifold. The $N=0$ level is equivalent to the lowest LL of a conventional system like GaAs, while the $N=1$ level is approximately equivalent to the conventional first LL.
In the situation of interest the electron spin is polarized by the Zeeman field,\cite{hunt2017direct} so we drop $\sigma$ in what follows, focusing on four components labeled by $\xi N$. 

\begin{figure}
\includegraphics[width=0.95\linewidth]{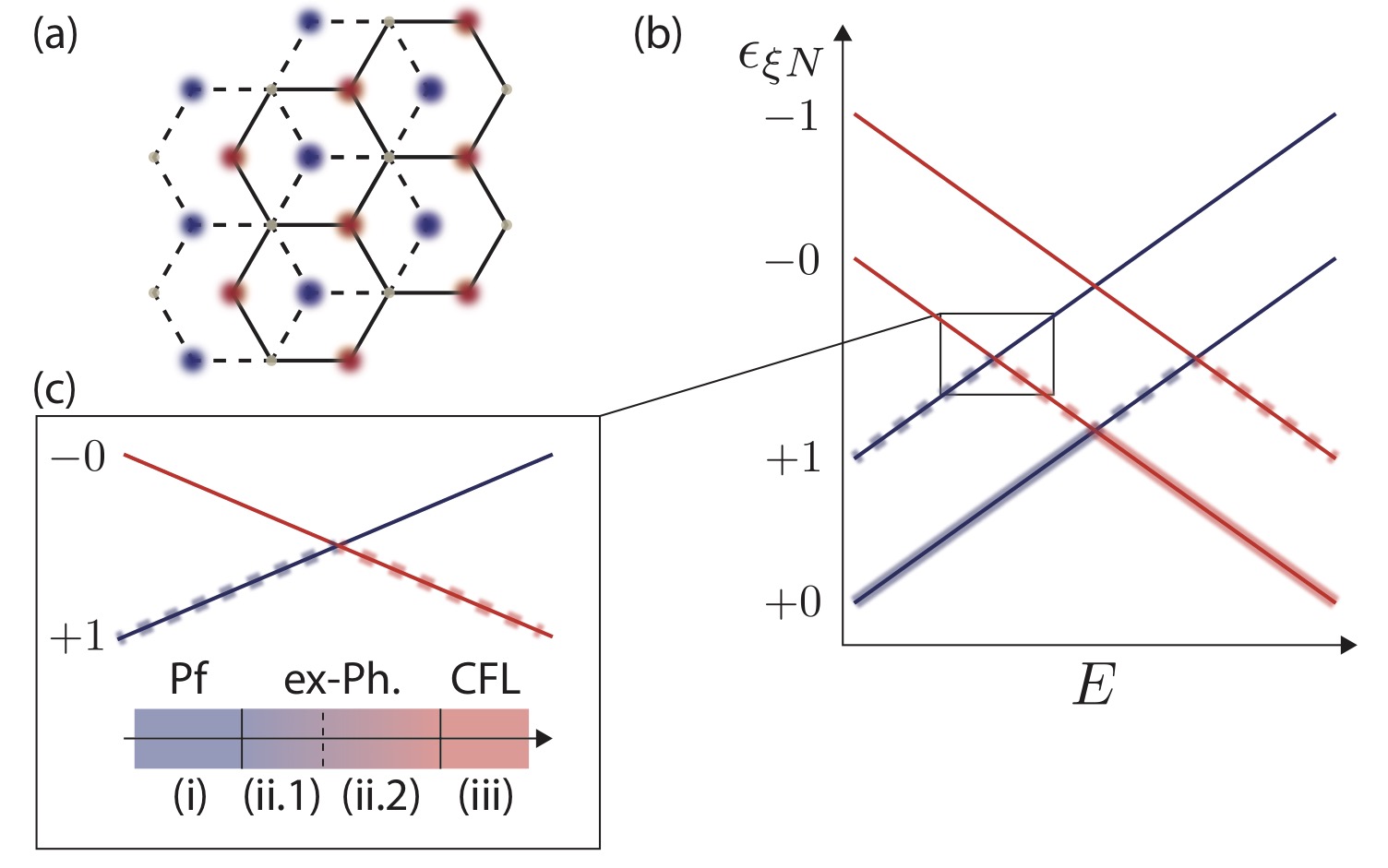}
\caption{(\textbf{a}) Bernal stacked bilayer graphene. In the zero-energy Landau level, electrons in valley $\xi = K_+$ are localized on sites of the bottom (red) layer, while in valley $K_-$, electrons are localized on the top (blue) layer. (\textbf{b}) Landau level energy spectrum $\epsilon_{\xi N}$, assuming spin polarization. An electric field $E$ across the bilayer acts as a valley Zeeman field, tuning level crossings. The situation of interest is at LL filling $\nu_T = 1 + \frac{1}{2}$;  $\nu=1$ of the filling is inert in +0, while $\nu = \frac{1}{2}$ is transferred from +1 to -0. (\textbf{c}) Detail of the crossing between a $+1$ and $-0$ level at half-filling. Four regions are observed: (i) incompressible, unpolarizable (ii.1) incompressible, polarizable (ii.2) compressible, polarizable (iii) compressible, unpolarizable.
In our theoretical proposal, these phases are identified as (i) the single-component Pfaffian order (ii.1) the exciton metal (ii.2) a layer-unpolarized two-component CFL (iii) the layer-polarized  single-component CFL. }
 \label{fig:BLG}
\end{figure}

In addition to this large LL degeneracy, a second interesting feature of BLG at finite-$B$ is that electrons in valley $\xi = +$ are localized onto the top layer of the BLG, while electrons in valley $\xi = -$ are localized onto the bottom layer, Fig.~\ref{fig:BLG}a. This feature is a peculiarity of the quadratic band touching, see the review Ref.~\onlinecite{McCannBLGReview}.
An electric field applied across the bilayer thus acts like a ``valley Zeeman'' field, Fig.~\ref{fig:BLG}b,  which can be used to establish a valley imbalance.
Since valley equals layer, an inter-valley exciton is simultaneously an inter-layer exciton;  however, the separation between the layers is tiny, $d \sim 0.35$nm, so it is  the mismatch in crystal momentum, $K_+ - K_-$, which prevents exciton relaxation.
These features combine to make BLG a novel platform for studying exciton phases: excitons are strongly bound due to the atomic scale inter-layer separation $d$, are long lived due to their crystal momentum $K_+ - K_-$, and carry a dipole moment $e d \hat{z}$ perpendicular to the layers which couples directly to an electric field or optical probes.

	As the perpendicular electric field $E$ is varied, the energies of the four relevant LLs cross as shown in Fig.~\ref{fig:BLG}(b). Previous theoretical studies have considered when $\nu_T$ is an integer; however, in this case there is no topological order, so the inter-layer excitons are necessarily of the familiar bosonic kind. Motivated by the recent experiments in BLG,\cite{Zibrov} we consider filling $\nu_T = 1 + \frac{1}{2}$ (measured with respect to an empty ZLL).
As illustrated in the LL spectrum of Fig.~\ref{fig:BLG}b, at large negative $E$ the electrons are polarized into valley (layer) $+$, which we write $\nu_{+0} = 1, \nu_{+1} = \frac{1}{2}$ [region (i) of  Fig.~\ref{fig:BLG}c].
Because electrons half-fill an $N=1$ LL, the situation is roughly analogous to the $\nu = \frac{5}{2}$ plateau of GaAs, and  experimentally a large ($\sim$ 1.8K) FQH gap is observed,\cite{Zibrov, li2017even}  consistent with the non-Abelian Pfaffian topological order \cite{MooreRead} we will describe in more detail shortly. As $E$ decreases, there is a  crossing between the $+1$ and $-0$ levels.
After the first crossing, the filling is  $\nu_{+0} = 1, \nu_{-0} = \frac{1}{2}$ [region (iii)]; since electrons half-fill a lowest LL, the situation is analogous to $\nu = \frac{1}{2}$ state of GaAs, and the system is compressible, consistent with the formation of a composite Fermi liquid (CFL). \cite{HLR}
The open question is the nature of the transition between them, at intermediate polarization $\nu_{+1} = \frac{1}{2} - \delta, \nu_{-0} = \delta$ [region (ii)].

Experimentally, Ref.~\onlinecite{Zibrov} observed that (1) there is a critical $E$-field at which the polarization $\delta$ begins evolving smoothly with the applied field, suggesting a continuous phase transition; (2) for small $0 < \delta \lessapprox 0.18$ [region (ii.1)] the system is incompressible but has finite polarizability (e.g., $\frac{d \, \delta}{dE} \neq 0$).
Since $\delta$ can be thought of as the density of inter-layer excitons, this suggests there is an intermediate charge-insulating phase of excitons; and (3) for intermediate $0.18 \lessapprox \delta < 0.5$ [region (ii.2)], the system is compressible and polarizable.

In our schematic, we have drawn the level crossing as un-avoided, which is true if charge is separately conserved in each valley.
In this case, the polarization $\delta$ is conserved, so can only change if the neutral gap closes. In other words, finite  polarizability $\frac{d \delta}{dE}$ implies that it costs infinitesimal energy to transfer a charge between the layers.
Due to their differing compressibility and polarizability, the four regions discussed above are then distinct phases of matter.
The most is intriguing is the nature of the charge insulating, but polarizable phase found for small $\delta$, region (ii.1).

As discussed here, the conservation of valley polarization is only protected by crystal symmetry, which one may worry isn't robust.
We first note that elsewhere in the BLG phase diagram an analogous $+0, -1$ level crossing exists in which the two components also have opposite spin, which further prevents tunneling since spin-orbit coupling is negligible, and the same phenomenology is observed.\cite{AndreaPersonalComm}
When the two components do have the same spin, short range disorder will manifest as dilute inter-layer hopping with a phase that is effectively random due to its dependence on the position of the impurity, $e^{i (K_- - K_+)\cdot R_{\textrm{imp}}}$.
In principle a 3-body umklapp term is also allowed,  which only conserves the relative charge modulo three, though this is expected to be  weak and suppressed by the ratio of the lattice-scale to magnetic length.
To assess the magnitude of these effects experimentally, Ref.~\onlinecite{Zibrov}  found that at filling $\nu_T = 1$ the crossing between the $+0$ and $-0$ levels leads to an extremely sharp transition where the polarizability spikes dramatically, which suggests these effects are very weak, since there would otherwise be a smooth, avoided crossing.
So, with or without the further $S^z$ protection, tunneling between the valleys appears negligible.
Regardless, in the exciton FS to be discussed the disorder scattering and umklapp are irrelevant in the RG sense.
For these reasons we will assume that charge is conserved separately in each layer.

\subsection{Hamiltonian}
Assuming the electrons in $\nu_{+0}=1$ are inert, the system is well approximated by a Coulomb interaction between the two components $-0, +1$:
\begin{align}
H =  \frac{1}{2}  \int d^2 \mathbf{q}  \sum_{i, j =0, 1} n_{i}(-\mathbf{q}) V_{ij}(q) n_{j}(\mathbf{q}) + \frac{E_V}{2} (\hat{N}_1 - \hat{N}_{0}) \label{eq:H}
\end{align}
Length is in units of the magnetic length $\ell_B$, and energy in units of the Coulomb scale $E_C = \frac{e^2 }{\epsilon \ell_B}$.
We neglect LL-mixing, so $n_{0}(\mathbf{q})$ is the Fourier-transformed density in component  $-0$, and $n_1(\mathbf{q})$ is the density in component $+1$. $E_V$ is the single-particle energy difference between the valleys, which (in the BLG zeroth-LL) is tuned by the electric field.

Neglecting the effect of screening and various valley anisotropies, the intra and inter-layer interaction is $V_{ii}(q) = \frac{2 \pi}{q}$, $V_{\textrm{01}}(q) = \frac{2 \pi}{q} e^{-q d}$ respectively, where $d \sim 0.05 \ell_B$ is the layer separation at $B=14$T. Since $d / \ell_B \ll 1$, we will set $d = 0$ unless specified otherwise; the other neglected valley-anisotropies are of comparable magnitude, and all of them  are suppressed by a factor of $d / \ell_B$ relative to the energy scale of interest. 

Even if the bare interaction $V_{ij}(q)$ is assumed to be SU(2)-symmetric, the \emph{effective} interactions between the two components will not be. This is because the two components are in different LLs, and when  density $n_i(\mathbf{q})$ is projected into LL $N=i$ it picks up a ``form factor'' $F_{ii}(q)$, with resulting effective interaction
\begin{align}
V^{\textrm{eff}}_{ij}(q) &= F_{ii}(q) V_{ij}(q)  F_{jj}(-q) \\
F_{00}(q) &= e^{-q^2 / 4}, \quad F_{11} = e^{-q^2 / 4}(1 - \frac{q^2}{2})
\label{eq:FF}
\end{align}
The $N=1$ form factor leads to a softer interaction. The large breaking of SU(2) by the different character  of the LLs is  why the other smaller valley anisotropies can be ignored.

\subsection{Theoretical proposal: the exciton metal}
	We briefly review the  proposal  of Ref.~\onlinecite{Barkeshli2016}.
We pass to the CF picture by attaching two-flux to electrons in \emph{both} components, leading to two species of CF.
At half-filling, the effective field seen by the CFs vanishes, $B_{\textrm{eff}} = (1 - 2 \nu) B = 0$.
For $\delta = 0$, all CFs reside in the $N=1$ LL, where the interactions are soft and favor pairing. The CFs  pair and form a spinless $p + i p$  superconductor - the ``Pfaffian'' phase.
There is compelling numerical evidence that the Pfaffian state is the ground state at half-filling of an $N=1$ LL.
Note that in bilayer graphene, ``Landau level mixing'' was theoretically shown to favor the Pfaffian over the anti-Pfaffian state,\cite{Zibrov} which turns out to be important for the energetics of the proposal of Ref.~\onlinecite{Barkeshli2016}.

To understand the excitons introduced at finite $\delta$, we review two of the relevant topological excitations of the Pfaffian phase.
A broken CF-Cooper pair generates a BdG quasiparticle $\psi_{\textrm{NF}}$, the neutral fermion,  an anyon which carries fermion parity but no electric charge.
The energy to create a neutral fermion is $\Delta_{\textrm{NF}} \sim 0.015 - 0.02 E_C$.\cite{FeiguinNF, MollerNF}
On the other hand, threading $4 \pi$ flux through the system generates a charge -$e$ bosonic excitation $\bqh$.
The elementary electron $\psi_1$ is a composite state of the two, $\psi_1 \sim \bqh \nf$.

Due to  the atomic scale proximity of the layers, when $\delta > 0$ the electrons in the $-0$ layer will bind to  holes in the $+1$ layer, both at density $\delta$ per flux.
But from the discussion above, there are actually \emph{two} types of excitons which are possible.
The conventional bosonic exciton is $\bex^\dagger = \psi^\dagger_0 \psi_1$, where $\psi_i$ is the electron operator in layer $i$. This is the familiar type of exciton whose signatures were detected in semiconductor QH bilayers at filling $\nu_T=1$~\cite{Spielman,tutuc2004counterflow}.
However, for energetic reasons it may be more favorable to bind the charge-$e$ boson, $\fex^\dagger = \psi_0^\dagger \bqh$, forming  a ``fermionic exciton.''
While this sounds exotic, Ref.~\onlinecite{Barkeshli2016} provided numerical evidence that it is the $\fex$ which has  lower energy in BLG at filling $\nu_T=1+\frac{1}{2}$, as we will soon explain.

Finite $\delta$ corresponds to a finite density of excitons on top of the Pfaffian phase, so if it is the $\fex$ which have lower energy they could form a Fermi surface.  
While the resulting exciton metal is charge insulating, it would have the thermal properties of a metal, metallic inter-layer counterflow, and Friedel oscillations at $2 k_F = 4 \delta \ell_B^{-1}$~\cite{Barkeshli2016}.

Note that in contrast to other neutral Fermi surface scenarios such as the spinon Fermi surface, which couple to an emergent gapless U(1) gauge field, here the fermionic excitons  couple to an emergent gapped $\mathbb{Z}_2$ gauge field (this is because the gauge field is Higgsed by the CF-superconductor).
For this reason, the exciton Fermi surface should be a Fermi liquid with linear-$T$ heat capacity.
However, as with any Fermi surface, it could also be unstable to localization by disorder, charge density order, or pairing.
In Ref.~\onlinecite{Barkeshli2016}, ED study has shown that a single $\fex$ is lower in energy than the $\bex$. In the following, we provide a microscopic picture for why this is the case. We then present further numerical evidence using ED and iDMRG for the stability and properties of the exciton phase containing a finite density of $\fex$.

\section{Microscopic picture for the stability of the fermionic exciton \label{sec:picture}}

\begin{figure}
\includegraphics[width=1.\linewidth]{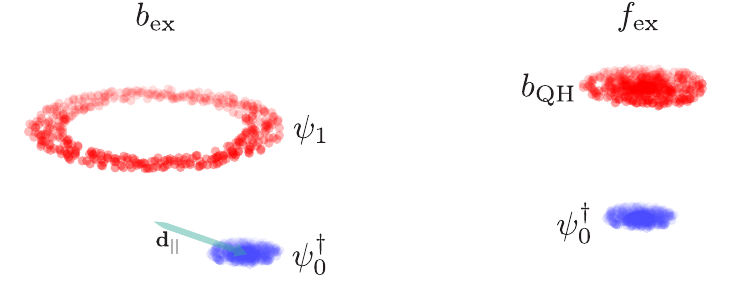}
\caption{The electrostatic consideration which determines the binding energy and dispersion-relation of an exciton. \textbf{Left}) In the $N=1$ LL, the electron $\psi_1$ can be visualized as a ring of radius $r \sim \ell_B$, due to the shape of the LL-orbit, while in an $N=0$ LL, the electron $\psi_0$ is a point. The conventional exciton is then frustrated by its in-plane dipole moment $\mathbf{d}_{\parallel}$.
\textbf{Right}) The fractionalized charge$-e$ boson $\bqh$ of the Pfaffian phase has a point-like charge distribution (see Fig.~\ref{fig:density}) despite existing in the $N=1$ LL. This gives the fermionic exciton $\fex$ a better binding energy.}
\label{fig:exblobs}
\end{figure}

It is instructive to warm up with an analysis of the exciton problem at integer filling, $\nu_T = 1$. We consider two layers which are in LLs $N=a$ and $N=b$ respectively, and starting from $\nu_b = 1, \nu_a = 0$ introduce an exciton.
The LL index $N$ changes the shape of the electron wavefunctions, and hence affects the binding energy of the exciton.
When $N=0$, an electron (or hole) inserted at the origin has a Gaussian profile  $n(r) \propto e^{-r^2/2}$, while in an $N=1$ level the wave-functions have a ring-like shape with $n(r) \propto r^2 e^{-r^2/2} / 2$.
If the hole is in an $N=1$ level, while the electron is in an $N=0$ level,  the binding energy of the exciton arises from the Coulomb attraction between a charge $-e$ point and a charge $e$ ring.
Clearly this binding energy will be less favorable than the point-point case, and furthermore, the attraction will be maximized when the point is displaced from the center of the ring, leading to a bound state with an intrinsic in-plane dipole moment $\mathbf{d}_\parallel \sim e \ell_B$.
This leads to the peculiar situation in which the exciton has an internal degree of freedom, its dipole moment, so that a  condensate would have to break rotational symmetry by choosing a dipole orientation. This degeneracy will  frustrate condensation.

The analysis can be made more quantitative by calculating the exciton's dispersion relation\cite{KunYang} analytically.
When ignoring LL-mixing, we can write down an exact exciton eigenstate of momentum $k$ and calculate its Coulomb energy $\epsilon(k)$,
\begin{align}
\epsilon_{ab}(k) &= \frac{1}{2 \pi} \int dq  \, q  V^{\textrm{eff}}_{ab}(q) (1 - J_0(q k) ), \\
 V^{\textrm{eff}}_{ab}(q) &= V_{\textrm{inter}}(q) e^{-q^2 / 2} L_a(q^2/2) L_b(q^2/2) 
\end{align}
where $V_{\textrm{inter}}$ is the interlayer Coulomb potential, $L_a$ is the Laguerre polynomial, and $J_0$ is the zeroth Bessel function. For $d=0$ interlayer separation, an exciton between two $N=0$ LLs has dispersion  $\epsilon_{00}(k) = \frac{1}{4} \sqrt{\frac{\pi}{2}} k^2 + \cdots$, with a unique minimum at $k = 0$ where the bosonic exciton can condense.
In contrast, between an $N=0$ and $N=1$ LL, $\epsilon_{01}(k) = -\frac{1}{8} \sqrt{\frac{\pi}{2}} k^2 + \frac{9}{128} \sqrt{\frac{\pi}{2}} k^4 + \cdots$ which has  a ``sombrero'' form with a degenerate minima that will strongly frustrate condensation.
The expressions are more involved for layer separation $d > 0$, but the sombrero-shape persists until $d \gtrapprox 0.8 \ell_B$.

This sombrero dispersion relation can be related back to the real-space  picture.
When ignoring LL-mixing, a neutral excitation of momentum $\textbf{p}$ has an in-plane dipole moment $\mathbf{d}_{\parallel} = \frac{e \ell_B^2}{\hbar} \hat{z} \times \mathbf{p}$.
For the exciton, $\mathbf{d}_{\parallel}$ is  the average displacement between the particle and the hole.
Since the ring-like nature of the $N=1$ hole prefers non-zero $\mathbf{d}_{\parallel}$, the exciton's dispersion relation has a minimum at $\mathbf{p} \neq 0$.

What does the $\nu_T=1$ analysis tell us about $\nu_T = \frac{1}{2}$? As discussed, the Pfaffian has two possible charge $e$ excitations: the electron-hole $\psi_1$, and the bosonic quasihole $\bqh$.
Following the integer discussion, the dispersion relation of an exciton formed from one of these holes and an electron $\psi^\dagger_0$ in the $N=0$  layer will depend on the charge distribution of the hole. We will show that creating an electron-hole $\psi_1$ on top of the Pfaffian background leads to the same ring-like shape as the $\nu_T = 1$ case, while the $\bqh$ takes the  form of a concentrated point, significantly lowering the energy of the $\fex$.

\begin{figure}
\includegraphics[width=0.95\linewidth]{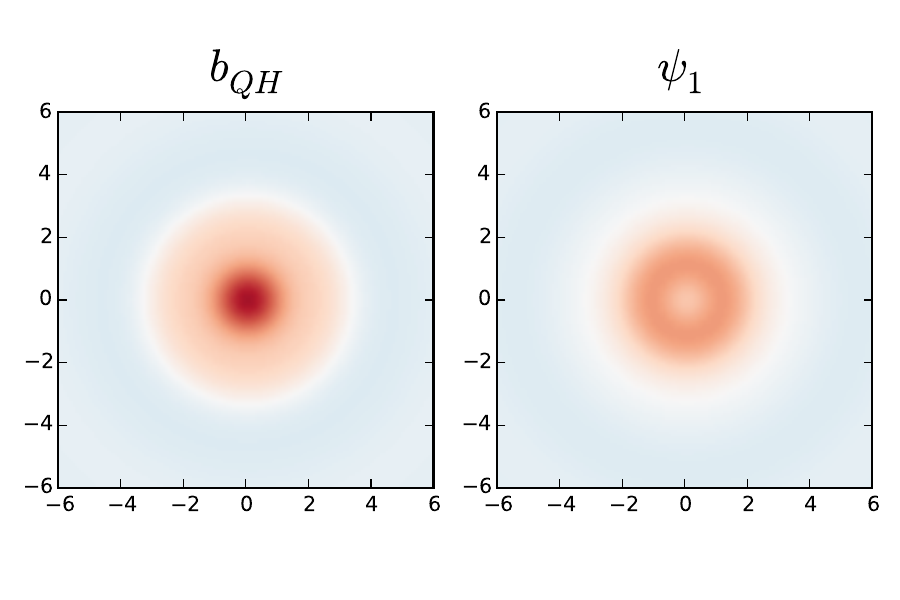}
\caption{Charge distribution of the bosonic quasihole  (labeled $\bqh$) and the electron-hole  (labeled $\psi_1$) of the Pfaffian state.
Calculations were performed on a sphere with the hole centered at the pole, and we show a color plot of the resulting electron density.
We see that the $\bqh$ density is centered on the pole, while $\psi_1$ forms a ring.
The calculations were done using the model 3-body interaction;\cite{MooreRead} $\bqh$ is obtained for $N_e = 14$ electrons in $N_{\textrm{orb}} = 28$ orbitals, while $\psi_1$ is obtained from $N_e = 13$ electrons in $N_{\textrm{orb}} = 26$ orbitals.}
\label{fig:density}
\end{figure}

	The shape of the quasiholes, and the resulting binding energy of the bosonic and fermionic excitons, can be determined analytically if we assume the Pfaffian phase is well described by the model wavefunction of Moore and Read.\cite{MooreRead}
Working in the symmetric gauge, we let $z_i$ run over the electron coordinates in the nearly full $N=1$ level, and $w_j$ the coordinates of the nearly empty $N = 0$ level.
For the purposes of presenting the wavefunction, we will temporarily pretend that the $z$ lie in an $N=0$ LL, so that the wavefunction is holomorphic in the symmetric gauge. The Pfaffian wavefunction is
\begin{align}
\Psi_{\textrm{Pf}}[\{z\}] &= \textrm{Pf}\left[ \frac{1}{z_i - z_j}\right]  \prod^{N_1}_{i < j} (z_i - z_j)^2,
\end{align}
where $N_1$ is even and we have ignored the usual Gaussian factor.\cite{Girvin} 
According to Moore and Read, a bosonic quasihole at $\eta$ is given by 
\begin{align}
\Psi_{\bqh}[\{z\}; \eta] = \textrm{Pf}\left[ \frac{1}{z_i - z_j}\right] \prod^{N_1}_i (z_i - \eta)^2 \prod^{N_1}_{i < j} (z_i - z_j)^2.
\end{align}
To form a fermionic exciton at momentum $k = 0$, we pin an electron $w_1$ to the location of the $\bqh$,
\begin{align}
\label{eq:Psifex}
\Psi_{\textrm{f-Ex}}[ \{ z\}, w_1 ] &=  \textrm{Pf}\left[ \frac{1}{z_i - z_j}\right]  \prod^{N_1}_{i} (w_1 - z_i)^2 \prod^{N_1}_{i < j} (z_i - z_j)^2,
\end{align}
and the total number of electrons $N_1 + 1$ is now odd.

In light of the $\nu_T = 1$ discussion, the key question is whether the electrons in the two layers efficiently avoid each other in real space. While each $z_i$ of $\Psi_{\textrm{f-Ex}}$ has a second-order zero at the location of $w_1$, the $z_i$  must be reinterpreted as $N=1$ LL wavefunctions.\cite{MorfCFL94}
Recall the single particle Hilbert space is  spanned by $\ket{N, n}$ for $N, n \geq 0$, where $N$ is the LL index and $n$ labels states within the LL.\cite{Girvin} The angular momenta of these states are $L^z = - \hbar (n - N)$.
In the LLL, $\phi_{N=0, n}(z) \propto z^n e^{-\frac{1}{4} |z|^2}$, which leads to the holomorphic form used above.
To obtain the actual wavefunction, however, we implicitly raise each $z$-particle from $\ket{N=0, n} \to \ket{N=1, n}$. 
We can determine the order of the correlation-hole without carrying out this promotion in full.
Fixing $w_1=0$, there is a good (first-order) interlayer correlation-hole if the $N=1$ particles are never at the origin.
While in the LLL only the $n=0$ orbital has weight at the origin, in the $N=1$ level it is the $n=1$ orbital which has weight at the origin, 
$\phi_{N=1, n=1}(z)  = (1 - |z|^2/2) e^{-\frac{1}{4} |z|^2}$ (more generally, the orbitals with $L^z = 0$ do, e.g. $n = N$).
Thus in the holomorphic language, for $w_1=0$ the $z$ should never have a \emph{first}-order zero at the origin;  the $(w_1 - z_i)^2$ term in the f-Ex wavefunction guarantees this constraint.
This argument can be verified by numerically calculating the density profile of the bosonic quasihole after doing the full promotion to the $N=1$ LL, as shown in Fig.~\ref{fig:density}a) - the electron density indeed has a first-order zero at the origin.

The wavefunction for the exciton metal was proposed to be \cite{Barkeshli2016}
\begin{align}
\nonumber \Psi_{\textrm{ex-Metal}}[ \{ z, w \} ] =  \mathcal{P}_{\textrm{LLL}} \textrm{Det}_{i, j}\left[ e^{ i ( \bar{k}_i w_j +  k_i \bar{w}_j)/2 } \right]  \textrm{Pf}\left[ \frac{1}{z_i - z_j}\right] \\
\times  \prod^{N_0}_{i < j} (w_i - w_j)^2 \prod^{N_0, N_1}_{i, j} (w_i - z_j)^2 \prod^{N_1}_{i < j} (z_i - z_j)^2, 
\end{align}
Here the $\textrm{Det}$ factor puts the $\{w\}$ into a Fermi sea at momenta $\{k\}$, entirely analogous to the Halperin-Lee-Read (HLR) wavefunction,\cite{RezayiReadCFL, HaldaneRezayi2000} and $\mathcal{P}_{\textrm{LLL}}$ projects into the $N=0$ LL.
Due to the $(z_i - w_j)^2$ factor, before projection each electron $w_j$ in the $N=0$ level is tied to a $\bqh$ in the $N=1$ level; thus placing the $w_j$ into a Fermi sea puts the $\fex$ into a Fermi sea.
It can be shown that projection into the LL shifts the zeros in proportion to $k$,\cite{read1996recent} $(w_i - z_j)^2 \to (w_i  + i k_i - z_j)^2$. As for the HLR state, this gives an $\fex$ at finite $\mathbf{k}$ a dipole moment $\mathbf{d}_{||} = e \ell_B^2  \hat{z} \times \mathbf{k} $  which costs Coulomb energy,  generating the ``kinetic energy'' term required for a robust Fermi surface.

In contrast, the bosonic exciton condensate at $k = 0$ has wavefunction \cite{Barkeshli2016}
\begin{align}
\Psi_{\textrm{ex-Cond.}}[ \{ z, w \} ] &=  \textrm{Pf}\left[ \frac{1}{x_i - x_j}\right] \prod_{i < j} (x_i - x_j)^2
\end{align}
where the $x_i$ run over both $z$ and $w$, and the $z$ are implicitly promoted to the $N=1$ LL.  For a single bosonic exciton we have 
\begin{align}
\label{eq:Psibex}
\Psi_{\textrm{b-Ex}}[ \{ z \}, w ] &=  \sum_k^{N_1} (-1)^k \frac{1}{w - z_k} \textrm{Pf}'\left[ \frac{1}{z_i - z_j}\right]  \\
 & \times \prod^{N_1}_{i < j} (z_i - z_j)^2 \prod^{N_1}_i (z_i - w)^2,\nonumber
\end{align}
where $\textrm{Pf}'$ denotes the Pfaffian factor in which coordinate $z_k$ is omitted, and $N_1 + 1$ is even.
As far as the $z$ are concerned, this is precisely the Moore-Read wavefunction with an electron hole $\psi_1(w)$ placed at $w$.\cite{MooreRead}
The key observation is that due to the $\frac{1}{w - z_k}$ term, there is always one particle $z_k$ which has only a first-order zero with respect to $w$.
When promoting to the $N=1$ LL, this implies that electrons on the two layers are sometimes coincident, increasing the interaction energy.
This  can be verified by numerically calculating the density profile of the electron-hole $\psi_1$,  shown in Fig.~\ref{fig:density}b), which forms a ring with non-zero density at the origin.

In summary, we see from the structure of the Pfaffian wavefunction that the point-like $\bqh$ is  much better suited for forming an exciton, which is the microscopic reason why $\fex$ is the lowest-energy exciton and has approximately quadratic dispersion relation.
The Coulomb interaction is expected to only quantitatively modify this picture, as confirmed by the lower exact diagonalization energy of the $\fex$  found in Ref.~\onlinecite{Barkeshli2016} and presented in further detail here.

\section{Exact diagonalization calculations: evidence for an exciton Fermi surface \label{sec:ed}}

\begin{figure}[t]
\centering
\includegraphics[width=0.5\textwidth]{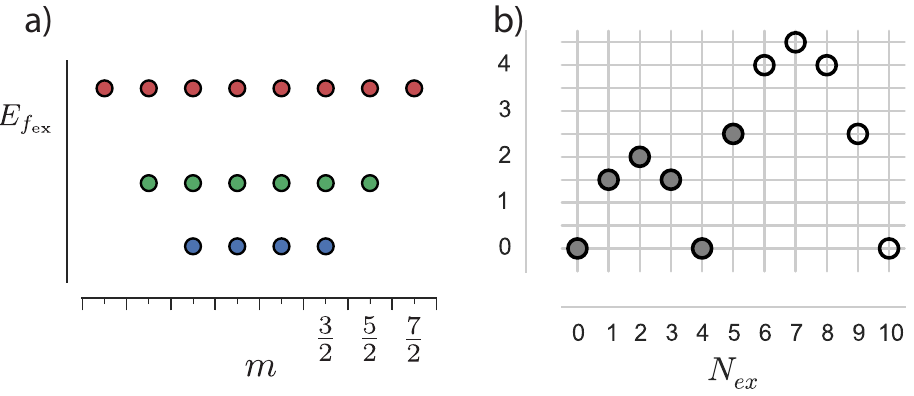}       
\caption{ \textbf{a}) The theoretically predicted single-$\fex$ energy levels. The states form degenerate multiplets according to their angular momentum $l$, here shown against the $z$-component $m$. Due to the $N_{\phi}^{\textrm{CF}} = 3$ flux-quanta seen by the CFs on the sphere, the energy levels should occur at $l = \frac{3}{2}, \frac{5}{2}, \cdots$. 
\textbf{b}) The predicted angular momentum $l$ of a Fermi sea containing $N_{ex}$ fermionic excitons. The $\fex$ sequentially fill the shells  starting with $l = \frac{3}{2}$. The angular momentum of a partially filled shell is determined by Hund's rule: in order to minimize the repulsive interactions, the $\fex$  maximize $l$ consistent with Pauli exclusion and antisymmetry. In contrast, if the interactions were attractive, the $N_{ex} = 2$ case, for instance, would have $l = 0$.
The filled circles indicate cases where we can obtain exact diagonalization data; Fig.~\ref{fig:ED} demonstrates the agreement with the predicted $l$.
 }
  \label{fig:shells}
\end{figure}

\subsection{Hund's rule predictions}
	Before going into great detail, we outline the theoretically expected behavior of an exciton FS. We exactly diagonalize Eq.~\eqref{eq:H} on a sphere, keeping the Hilbert space of both an $N=0$ and $N=1$ LL. Since  charge is conserved separately in each LL, the spectrum can be diagonalized in sectors of fixed particle number $N_1, N_0$ for each LL.
For $N_e = N_1 + N_0$ total electrons, we start with the Pfaffian ($N_1 = N_e, N_0 = 0$), and study the energy spectrum as we add a small number $N_{ex}$ of excitons ($N_1 = N_e - N_{ex}, N_0 = N_{ex}$). 
The formation of a stable Fermi sea can be detected analogously to earlier exact diagonalization studies of the CFL.\cite{RezayiReadCFL}
If a FS forms, then at low energies the $\fex$ will be governed by an effective Hamiltonian of the form
\begin{align}
H_{\textrm{eff}} = \sum_{j=1}^{N_{ex}} \frac{\textbf{p}_j^2}{2 m_{ex}} + V_{\textrm{int}}.
\end{align}
The first term is a kinetic energy (which ultimately has its origin in the Coulomb interaction), and the second a residual effective interaction. Crucial for the formation of a FS is that $V_{\textrm{int}}$ be repulsive, otherwise the FS will be unstable.
On the sphere, the kinetic term becomes $\sum_j \frac{\mathbf{L}_j^2}{2 m_{ex} R^2}$, where $\mathbf{L}$ are the angular momentum operators of the $\fex$ and $R$ is the radius of the sphere.
The ``single particle'' states then come in degenerate multiplets according to $L^2$, like the shells of an atom (see Fig.~\ref{fig:shells}a). 
This leads to a characteristic evolution of the angular momentum $l$ of the lowest energy state as the $\fex$ are added, e.g. $l = 0$ whenever the outer shell is filled, and $l > 0$ otherwise (Fig.~\ref{fig:shells}b).
For partially filled shells, the degeneracy is lifted by $V_{\textrm{int}}$, which (if repulsive) will lead to a Hund's-rule by which the lowest energy configuration maximizes  $l$. Together these effects can confirm the existence of both the kinetic energy and a repulsive interaction.

One predicted peculiarity of the $\fex$ is that its  shells do  \emph{not}  carry the familiar angular momenta $l = 0, 1, \cdots$ of $s, p, d, \cdots$ orbitals.
Due to flux attachment, the number of flux quanta experienced by each CF (and hence the $\fex$) is
\begin{align}
N^{\textrm{CF}}_{\phi} &= N_\phi - 2 (N_e - 1) = 2 N_e  - 5 - 2 (N_e - 1) = 3
\end{align}
The sub-extensive effective magnetic field modifies the spectrum of the kinetic energy,\cite{Haldane83} $L^2 \in [ l (l + 1) - \left( \frac{3}{2} \right)^2 ] \hbar^2$, where $l = \frac{3}{2}, \frac{5}{2}, \cdots$, with  degeneracies $2 l + 1$. Thus we predict  that  the $\fex$'s shells instead begin with $l = \frac{3}{2}$, as shown in Fig.~\ref{fig:shells}.

\subsection{Single excitons}
	We now detail the numerical calculations. The Hilbert space is not  entirely analogous to a two-component spin system, because the $N=1$ LL  contains  two more orbitals than the $N=0$ LL at a given system size. Correctly accounting for this difference, rather than treating the $N=1$ LL as an $N=0$ LL with a modified interaction, is  crucial for observing the correct behavior. 
 In contrast to Ref.~\onlinecite{Barkeshli2016} where the bare Coulomb interaction was used, here	we add a small component  of the $m=1$ Haldane pseudopotential ($0.05 V_1$ in units of the Coulomb scale $E_C$) to the interactions within the $N=1$ level.
This small perturbation is required to stabilize the Pfaffian phase~\cite{HaldaneRezayi2000} and reduce finite-size effects, which is important when studying multiple $\fex$ excitons. (In the BLG experiments, Landau level mixing is believed to stabilize the phase~\cite{RezayiSimon, Zaletel2015, Zibrov}).
The Pfaffian ground state occurs when the number of magnetic flux quanta piercing the sphere satisfies $ N_\phi = 2 N_e  - 5$ (note that the ``shift'' \cite{Haldane83, WenZee} is $\mathcal{S} = 5$, rather than the shift $\mathcal{S} = 3$ usually associated with the Pfaffian, because we treat the $N_1$ particles as living in a $N=1$ LL).
Working at $N_\phi = 2 N_e  - 5$ throughout, we obtain the low lying energy spectrum $E_i(N_e, N_{ex})$, where $i = 0, 1, \cdots$ labels the energy levels, which come in degenerate multiplets according to their angular momentum.
All energies are quoted in units of a finite-size rescaled Coulomb energy $E'_C = \sqrt{\frac{2 N_e}{2 N_e - 5}} E_C$~\cite{Morf86} and the radius of the sphere is defined as $R=\sqrt{N_\phi/2}$.	

\begin{figure}[t]
\centering
\includegraphics[width=0.23\textwidth]{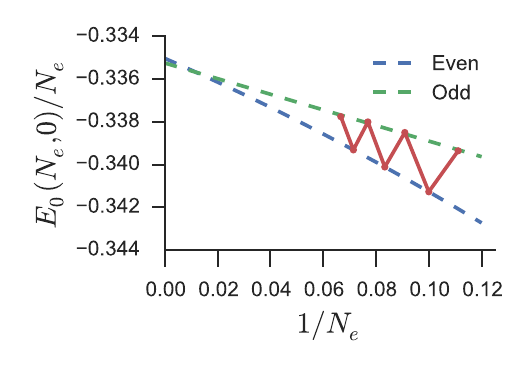}       
\includegraphics[width=0.23\textwidth]{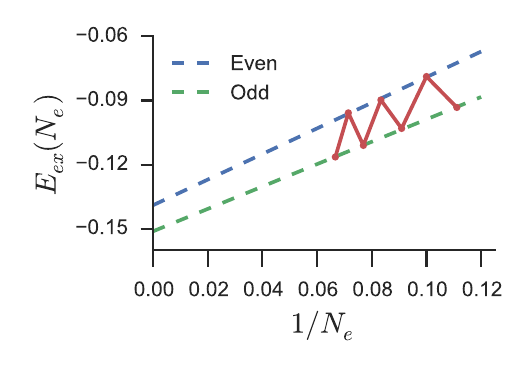}
\caption{\textbf{Left})  \label{fig:Egs}The ground state energy per electron with no excitons,  $E_0(N_e, 0) / N_e$ (solid red).
For $N_e$ even, we extrapolate $E_0(N_e, 0) \approx   E_{\mathds{1}}(N_e) \equiv  E_{\mathds{1}} N_e + a_{\mathds{1}} + b_{\mathds{1}} N_e^{-1}$ (dashed  blue) to obtain the vacuum energy per electron $E_{\mathds{1}}  = -0.3350$.
For $N_e$ odd we fit $E_0(N_e, 0) \approx   E_{\textrm{NF}} N_e + a_{\textrm{NF}}$ (dashed green). 
The extensive parts agree ($E_{\mathds{1}} = E_{\textrm{NF}} \pm 10^{-4}$), while the constant correction provides an estimate of the neutral fermion gap, $\Delta_{\textrm{NF}}  \approx a_{\textrm{NF}} - a_{\mathds{1}} = 0.017$.
\textbf{Right})  \label{fig:Eex} The energy of a single exciton, $E_{ex}(N_e) \equiv E_0(N_e, 1) - E_{\mathds{1}}(N_e)$, where we subtract off the second-order extrapolation of the vacuum energy. $E_{ex}(N_e)$ also shows an odd-even effect, but reversed: odd-$N_e$ is lower. Odd-$N_e$ corresponds to the $\fex$, while even the $\bex$.
  }
\end{figure}

	We first repeat the standard analysis of the Pfaffian phase for $N_{ex} = 0$.
Since the number of CFs is equal to $N_e$, when $N_{e}$ is even CF can pair into a CF-superconductor with a unique ground state (we call $N_{e}$ even, $N_{ex} = 0$ the ``vacuum'' sector).
When $N_e$ is odd, one CF remains unpaired, and this broken Cooper pair is precisely the $\nf$-excitation. 
Thus the odd-$N_e$ sector can be thought of as an excited state, with an energy which should be higher by the neutral-fermion gap $\Delta_{\textrm{NF}}$.
Fig.~\ref{fig:Egs}a shows that $E_0(N_e, 0)$ indeed displays an odd-even energy difference,  which we extrapolate in $1/N_e$ to estimate  $\Delta_{\textrm{NF}} = 0.017$, in line with earlier estimates.\cite{FeiguinNF, MollerNF}

	We introduce an exciton by studying the ground state energy $E_0(N_e, N_{ex} = 1)$.
To estimate the energy of the exciton $E_{ex}$, we subtract off a smooth \emph{extrapolation} of the vacuum energy, $E_{ex}(N_e) \equiv E_0(N_e, N_{ex} = 1) -  E_{\mathds{1}}(N_e)$, as shown in Fig.~\ref{fig:Eex}b. 
It is important to subtract off a smooth extrapolation $E_{\mathds{1}}(N_e)$, not the actual $E_0(N_e, 0)$, otherwise the odd-$N_e$ case will include an undesired subtraction of $\Delta_{\textrm{NF}}$.
There is again a characteristic odd-even effect, but reversed: \emph{odd} $N_e$ has lower energy. 
From the structure of Eq.~\eqref{eq:Psifex} and \eqref{eq:Psibex}, we see that the $\bex$ occurs for $N_e$-even, $N_{ex}=1$, while the $\fex$ occurs for $N_e$-odd, $N_{ex}=1$, since the $\fex$ eats up the unpaired CF.
Thus the odd-even energy difference is a smoking-gun signature that the $\fex$ is lower in energy than the $\bex$.
Separately extrapolating $E_{ex}(N_e)$ in powers of $1/N_e$ for both odd and even $N_e$, we find the energy difference is $\Delta_{\bex} - \Delta_{\fex} \sim 0.014$. 
This difference is roughly consistent with $\Delta_{\textrm{NF}} = 0.017$, which is expected since the unstable $\bex$ will fractionalize into a $\nf$ and an $\fex$.
	
	We now analyze the single $\fex$ spectrum in greater detail. In Fig.~\ref{fig:ED}, we indicate the $l$-values of the ground state for various $N_e, N_{ex}$. We find that for all $N_{ex} = 1$, $N_e$-odd (the $\fex$ sector), the ground state is an $l = \frac{3}{2}$ multiplet as predicted. 
The reader will notice that this is in contrast to the case of a single $\nf$ ($N_{ex} = 0$, $N_e$-odd), where we find $l > \frac{3}{2}$. The discrepancy arises because the band minima of the neutral fermion is at the Fermi wave vector $k_F = \ell_B^{-1}$ of the CFL: equating $k_F = l / R$,  $l$ should be the half-integer nearest to $\sqrt{N_e - 5/2}$, precisely as observed.

\begin{figure}[t]
\centering
\includegraphics[width=0.4\textwidth]{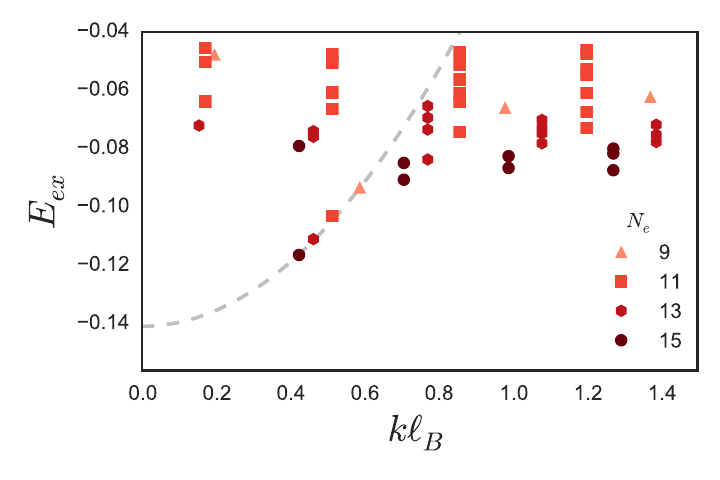}       
\caption{ The $\fex$ excitation spectrum $E_{ex}(N_e) \equiv E_j(N_e, 1) - E_{\mathds{1}}(N_e)$, $N_e$-odd. To collapse the data, we convert the angular momentum $l_j$ of a multiplet to a momentum $k_j = l_j / R(N_e)$, where $R$ is the radius of the sphere. The mode dispersing below the continuum is the putative $\fex$. A continuum is expected within $2 \Delta_{\textrm{NF}} \sim 0.04 E_C$ of the band minima due to neutral-fermion pairs.
Given the limited range of $k$, we do not take the quadratic fit to $\Delta_{\fex} + \frac{k^2}{2 m_{\fex}}$ (shown dashed) very seriously, but obtain $\frac{\ell_B^{-2}}{2 m_{\fex}} \sim 0.137 E_C$. This is a huge energy by FQH standards, indicating a very strong binding energy for the fermionic excitons. \label{fig:Eex_spec}}
\end{figure}

	In Fig.~\ref{fig:Eex_spec}, we show the low-lying $\fex$ excitation spectrum $E_j(N_e = \textrm{odd}, N_{ex}=1)$ as a function of the angular momentum $l$, collapsed across system sizes using $k = l/R$. We see an isolated branch  which merges into a continuum at $k \sim 0.6 \ell_B^{-1}$.  It is intriguing to note that in the experiment of Zibrov et al., \cite{Zibrov} the charge gap closes at $\delta = 0.18$, which corresponds to Fermi wavevector $\ell_B k_F = \sqrt{2 \delta} = 0.6$ -- right where we find the $\fex$ mode hits the continuum.
	
While the finite system size limits the lowest $k$ we can achieve (note the dimension of the $(15, 1)$ Hilbert space is over 561 million),  from a quadratic fit to the isolated branch we obtain a rough estimate of the $\fex$ mass $\frac{\ell_B^{-2}}{2 m_{\fex}} \sim 0.137 E_C$. While it should be taken with a grain of salt, using $k_F= \ell_B^{-1} \sqrt{2 \delta}$ this gives a Fermi energy of $E_F =  0.27 \delta \, E_C \sim 8 \delta \, \textrm{meV} \sim 101 \delta\, $K at $B = 14$T, $\epsilon = 6.6 \epsilon_0$. This would imply experiments at $\delta = 0.1$ could easily achieve the Fermi-degenerate regime. Furthermore, since the disorder width $W$ is more likely on the scale of 1K or less, the exciton-FS would appear delocalized above a vanishingly small crossover temperature $T^\ast \sim e^{-E_F / W}$.

\subsection{Multiple excitons: emergence of FS}

\begin{figure}[t]
\centering
\includegraphics[width=0.5\textwidth]{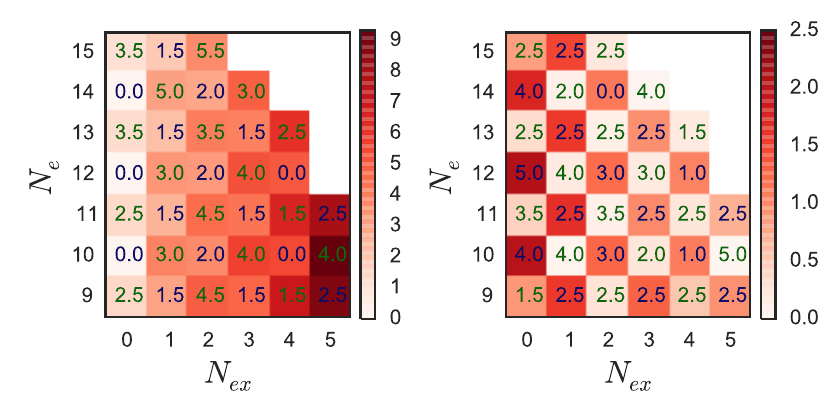}       
\caption{\label{fig:ED} \textbf{Left}) The ground state energy $E_{\textrm{FS}}(N_e, N_{ex}) \equiv E_0(N_e, N_{ex}) - E_{\mathds{1}}(N_e) - \Delta_{\fex} N_{ex}$ in the plane of $(N_{ex}, N_e)$ (see main text for an explanation of the subtraction). The colorbar indicates the energy in units of $\Delta_{\textrm{NF}} = 0.017$. The inset text gives the angular momentum $l$ of the ground state; blue for sectors predicted by the exciton-FS ($N_e + N_{ex}$ even), and green otherwise. In all exciton-FS sectors, the observed $l$ agrees with the Hund's rule prediction of Fig.~\ref{fig:shells}b.
\textbf{Right}) The neutral gap $E_{1}(N_e, N_{ex}) - E_{0}(N_e, N_{ex})$ between the two lowest multiplets, with colorbar in units of $\Delta_{\textrm{NF}}$. The inset text gives the $l$ of the first-excited multiplet. The exciton-FS sectors have a consistently larger gap, concomitant with the binding energy of the $\fex$ and repulsive $V_{\textrm{int}}$. Note that for a single $\fex$ ($N_{ex} = 1$, $N_e$-odd), the first excited state always has $l = \frac{5}{2}$; this is presumably the second shell of Fig.~\ref{fig:shells}a.
 }
\end{figure}

We test for the formation of an exciton FS by adding multiple excitons,  Fig.~\ref{fig:ED}. To summarize the data within a single figure, we define the energy of the FS using two subtractions: $E_{\textrm{FS}}(N_e, N_{ex}) \equiv E_0(N_e, N_{ex}) - E_{\mathds{1}}(N_e) - \Delta_{\fex} N_{ex}$. The first term removes the smooth part of the vacuum energy (Fig.~\ref{fig:Egs}a), and the second  is a shift which sets the chemical potential of the $\fex$ to zero.
Note that we are free to add $N_1 - N_0$ to the Hamiltonian without changing the eigenstates (indeed, this is the bias potential $E_V$ which tunes the transition, Eq.~\eqref{eq:H}), which shifts the energy spectrum in proportion to $N_{ex}$. We have chosen to shift by the  $\fex$ energy $\Delta_{\fex}$ estimated in Fig.~\ref{fig:Eex}b, which conveniently brings $E_{\textrm{FS}}$ within a narrow range of energies.

	The results are completely consistent with an exciton-FS across all $N_e, N_{ex}$,  see Fig.~\ref{fig:ED}(left). First, fixing $N_{ex}$, we always observe an odd-even effect in $N_e$ with the $(N_{ex} + N_e)$-even case having lower energy.
This is again smoking gun evidence in favor of fermionic excitons; bosonic excitons always occur for $N_e$-even.
Second, in the $(N_{ex} + N_e)$-even sectors, the angular momentum $l$ of the ground state always increases with $N_{ex}$ according to $l = 0, \frac{3}{2}, 2, \frac{3}{2}, 0, \frac{5}{2}$, in precise  agreement with the Hund's rule prediction of Fig.~\ref{fig:shells}b  (we are unable to go beyond $N_{ex} = 5$). The $N_{ex} = 2$ case is particularly non-trivial, because two $l = \frac{3}{2}$ fermions could fuse to either $l = 0, 2$. The preference for large relative angular momentum is an indication of repulsive interactions between the $\fex$, and hence stability against pairing. It would certainly be useful to verify that this repulsion persists for, e.g., $N_e = 16, N_{ex} = 6$, but such calculations are prohibitive. 

It is worth contrasting these observations with the expected properties of a bosonic exciton condensate, Eq.~\eqref{eq:Psibex}. First, the bosonic condensate would always occur for $N_{e}$-even, with $N_{e}$-odd higher in energy by $\Delta_{\textrm{NF}}$, counter to our findings. Second, in the case $N_{ex} = 2, N_e$-even, where the both the exciton-FS and bosonic condensate can occur, we find $l = 2$, while a condensate at $k = 0$ would presumably always have $l = 0$.

	In Fig.~\ref{fig:ED}(right) we show the gap to the first excited multiplet, $E_1(N_e, N_{ex}) - E_0(N_e, N_{ex})$, as well as its $l$. The $(N_{ex} + N_e)$-even sectors have consistently larger gaps.
Within the exciton-FS scenario, this is because the $(N_{ex} + N_e)$-odd sectors  contain an extra $\nf$ in a dispersing continuum.
Of course, in the thermodynamic limit  the $(N_{ex} + N_e)$-even gaps should go to zero as the single-exciton level spacing decreases and the particle-hole excitations decrease in energy. We do see the gaps decrease, but no definitive extrapolation can be made.

\subsection{Charge gap}

\begin{figure}[t]
\centering
\includegraphics[width=0.5\textwidth]{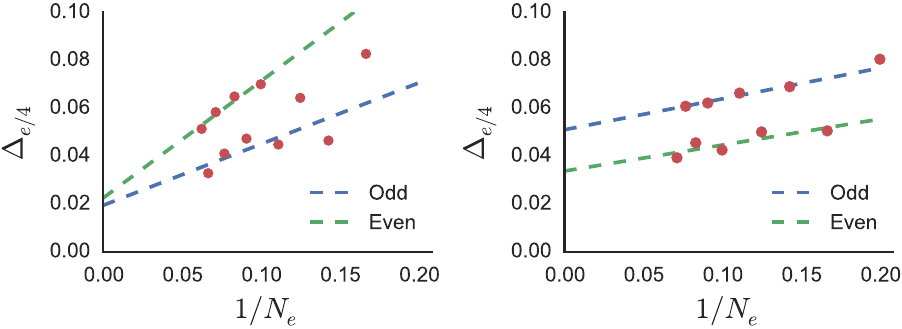}       
\caption{\textbf{Left}) The charge gap $\Delta_{e/4}(N_e)$ in the absence of excitons, $N_{ex} = 0$. We ``fit'' the last four data points, separately for $N_e$-odd, even, to estimate $\Delta_{e/4} \sim 0.02$. 
\textbf{Right}) The charge gap $\Delta_{e/4}(N_e)$ with the addition of a single exciton, $N_{ex} = 1$.  If anything, the charge gap is even larger. The fits in $1/N_e$ should not be taken too seriously: the true scaling behavior of these gaps is presumably very complicated, since they are three or four-quasiparticle states (two $\frac{e}{4}$s and an a $\fex$, or two $\frac{e}{4}$s, an $\fex$, and a $\nf$ for $N_{e}$-odd, even respectively).\label{fig:cgap}}
\end{figure}

Finally, in Fig.~\ref{fig:cgap} we consider the charge gap, which should remain finite. Since the exciton metal supports charge $\pm \frac{e}{4}$ excitations, the charge gap $\Delta_{e/4}$ is conventionally defined as the energy required to separate a charge  $e/4$, $-e/4$ pair.
Ideally, we would calculate this gap as a function of the polarization density $N_{ex} / N_e$ and extrapolate to $N_e \to \infty$; unfortunately, on the small grid of sizes available to us it is impossible to obtain two data points at the same polarization density.
So we resign ourselves to computing the charge gap in the presence of a single exciton, which is at least a consistency check.

	In the Pfaffian state, adding  one flux to the system nucleates two $\frac{e}{4}$ quasiparticles, at energy cost $E_+(N_e) = E(N_e, N_\phi + 1) - E(N_e, N_\phi)$. However, as discussed in detail by Morf~\cite{Morf02}, these energies contain large $\frac{1}{\sqrt{N_e}}$ scaling corrections due to the long-range part of the Coulomb interaction. To correct for them, we follow  the subtraction scheme discussed therein, the only difference being that in our case one particle is demoted to the $N=0$ LL:
\begin{align}
\Delta_{e/4} &= \frac{1}{2} \left[\tilde{E}_+  - 2 E_0 + \tilde{E}_- \right], \\
\tilde{E}_{\pm}(N_e) &= E(N_e, N_{ex} = 1, N_\phi \pm 1) + \frac{5}{32} \frac{1}{R_\pm} \\
E_{0}(N_e) &= E(N_e, N_{ex} = 1, N_\phi), \;\;\; N_\phi = 2 N_e - 5
\end{align}
where $R_\pm = \sqrt{(N_\phi \pm 1) / 2 }$.
Note the factor of $\frac{1}{2}$ in the definition of $\Delta_{e/4}$ arises because $N_\phi \pm 1$ nucleates \emph{two} quasiparticles. The resulting gaps are shown in Fig.~\ref{fig:cgap}, which  indeed remain finite.

In summary, exact diagonalization finds perfect agreement with the Hund's rule predicted by the formation of an exciton Fermi surface, and in sharp contrast to the expected behavior of a bosonic exciton condensate.

\section{iDMRG calculations \label{sec:dmrg}}

We next use iDMRG to calculate ground-state properties at finite $\nu_1 = \frac{1}{2} - \delta, \nu_0 = \delta$.
The iDMRG technique  has been established as an effective method for finding the ground state of a variety of quantum Hall systems\cite{Zaletel2013}, including for multicomponent systems \cite{Zaletel2015} and the gapless CFL state at filling $\nu=1/2$\cite{Geraedts}. 
The computational difficulty of the DMRG increases with the amount of quantum entanglement in the system, making the problem at hand extremely challenging. 
Capturing the single-component, gapped Pfaffian phase already requires significant resources (i.e., DMRG bond dimension $\chi \sim 6000$),\cite{Zaletel2015} while the gapless, single-component CFL required $\chi \sim 8000$ to manifest good scaling properties.\cite{Geraedts}
We are proposing to simulate a Fermi surface and Pfaffian phase \emph{together}. In some very crude sense the difficulty of DMRG is ``multiplicative'' when adding together degrees of freedom, so the problem is difficult indeed.

	The iDMRG supports (or is at least consistent with!) four claims: (1) there is a continuous $E_V$ tuned transition; (2) the finite $\delta$ state is a liquid with no signs of crystalline order (e.g., stripes or bubbles); (3) the polarization sector is gapless; (4) there is no evidence for off-diagonal long range order of a bosonic exciton.
Unfortunately, numerical limitations have frustrated our ability to directly characterize the putative exciton Fermi surface using entanglement measures or Friedel oscillations (see Appendix), so the DMRG cannot explicitly confirm that an exciton FS has formed.
While the evidence from iDMRG is somewhat more indirect than from exact diagonalization, it can reach much larger systems sizes, so the two approaches are nicely  complementary in this respect.

The iDMRG algorithm proceeds by placing the quantum Hall problem of Eq.~\eqref{eq:H} on an infinitely long-cylinder of circumference $L$; in this work, $L =  16 \ell_B$.
To make the Coulomb interaction well defined on the cylinder, for the iDMRG results we use a screened Coulomb interaction $V_{ij}(q) = \frac{2 \pi}{q} \tanh(D q)$, with $D \sim 8 \ell_B$, as is actually the case in BLG heterostructures. Since $D$ is  large compared to $\ell_B$, it is not expected to significantly alter the energetics. \cite{Zaletel2015}
To best stabilize the Pfaffian order, we add a small short-range component to the interactions within the $N=1$ LL  which has a similar effect as the $0.05 V_1$ perturbation used in ED (see Appendix).

The Hamiltonian conserves charge separately in each valley, so it is most convenient to set the splitting $E_V = 0$ and use iDMRG to find the ground-state $\ket{\delta}$ at fixed $\delta$. 
We know that when $\delta=0$ the system is in a Pfaffian phase, while when $\delta=1/2$ it is a CFL, and are interested in what happens when $\delta$ is between these two values. There are a number of possibilities which we can evaluate using iDMRG results.

\subsection{Evidence for a continuous transition: ground state energy}

\begin{figure}
\centering
\includegraphics[width=0.7\linewidth]{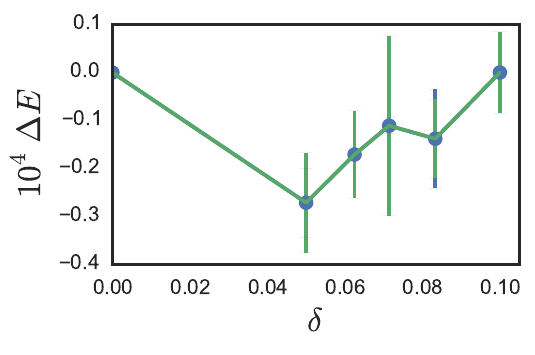}
\caption{Energy per flux $E(\delta)$ vs. $\delta$ for $L=16$, in units of $E_C$. We subtract a linear background, $\Delta E = E(\delta) - 10 \delta (E(\frac{1}{10}) - E(0))$. Within our error bars, $E(\delta)$  is concave up, consistent with a continuous phase transition between the Pfaffian ($\delta = 0$) and a CFL ($\delta = \frac{1}{2}$).
Energies were obtained by a fitting data at multiple DMRG bond dimensions $\chi$  to the form $E(\chi)=E_0+a\chi^{-b}$, with $\chi\in[5400,18000]$, with error bars taken from the discrepancy in the interpolation with and without the last data point.
While we would like to fill in the curve for smaller $\delta$,  iDMRG calculations at $\delta = \frac{p}{q}$  require a unit cell of length $q$, which prevents us from studying below $\delta = \frac{1}{20}$.}
\label{fig:energy}
\end{figure}

The first possibility is a first-order transition at which $\delta(E_V)$ jumps discontinuously at some critical value of the applied splitting $E_V$.
We cannot test this directly in our numerics since iDMRG forces a fixed, spatially uniform polarization $\delta$.
However, we can measure  the energy per flux quantum, $E(\delta)$, and so long as $E(\delta)$ is  convex-up ($\frac{d^2 E}{d \delta^2} > 0$), the polarization $\delta(E_V)$ is then determined by  Legendre transformation, $\frac{dE}{d \delta} = E_V$.
The convex-up scenario thus indicates a continuous transition. 
However, if we find $E(\delta)$ is concave-down ($\frac{d^2 E}{d \delta^2} < 0$), then states with uniform $\delta$ will have higher energy than those with phase-separation (by the Maxwell construction), indicating a first order transition.
Fig.~\ref{fig:energy} shows our results for $E(\delta)$:  while somewhat noisy, within the error bars of our numerics the data is concave up, consistent with a continuous transition.

We note that by taking the layer separation $d=0$, we are considering the scenario most likely to phase separate, since finite $d$ leads to an \emph{additional} concave-up capacitive charging energy. In fact, this capacitive energy is always  sufficient to prevent macroscopic phase separation. \cite{Jamei2005}
For uniform $\delta$, this capacitive energy is  $E^{\textbf{c}} = E_C \frac{d}{\ell_B} \frac{\epsilon_{\parallel}}{\epsilon_{\perp}} \delta^2$, where $\epsilon_{\parallel} / \epsilon_{\perp}$ is the ratio of the in-plane and perpendicular dielectric constants.  Based on measurements of BLG,  \cite{hunt2017direct} the capacitive contribution happens to be about the same order of magnitude as the curvature in $E(\delta)$, and hence would be important for quantitatively predicting $\delta(E_V)$.

 \begin{figure}
\includegraphics[width=0.95\linewidth]{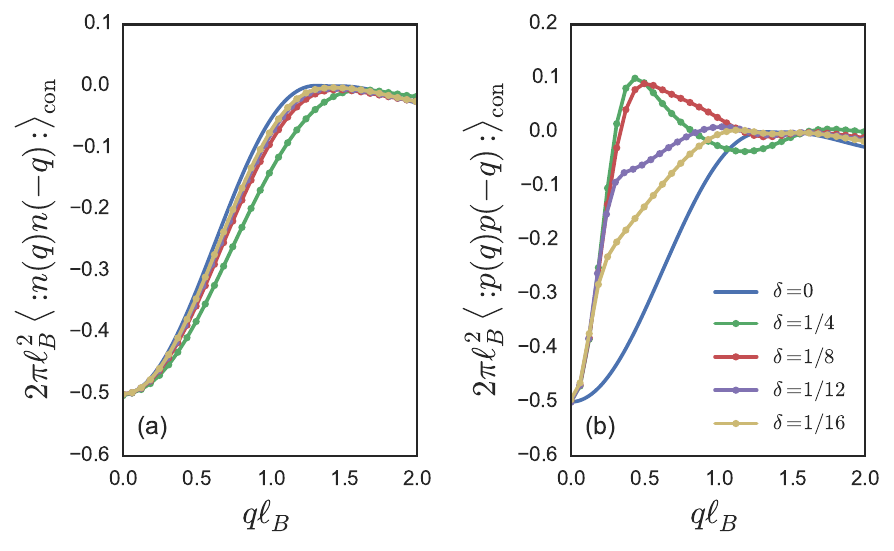}
\caption{({\bf a}) Charge-charge $S_{nn}(q)$ and ({\bf b}) polarization-polarization $S_{pp}(q)$ structure factors for a variety of  $\delta$.  Data was obtained for $L=16$ with $\chi=18000$. For both cases, there are no  sharp delta-function like  peaks, as would have occurred if there was a tendency towards a charge density wave order (i.e., stripes, bubbles, or other ``microemulsions''~\cite{Jamei2005}).
However, in contrast to the charge sector, which changes smoothly with $\delta$, the polarization-polarization data develops a non-analyticity at $q \to 0$ ($S_{pp}(q) \propto |q|$, cf. Fig.~\ref{fig:scaling}). This indicates the emergence of a gapless neutral mode at finite $\delta$. }
\label{fig:structure}
\end{figure}

 \subsection{Evidence for a liquid: structure factors}
 
 	A second possibility is some sort of stripe or bubble, either in the total charge or valley polarization. To asses this possibility, we examine correlation functions of $n = n_1 + n_0$ (the total density) and $p = n_1 - n_0$ (the  polarization). In the 2D limit,  density-wave order would manifest as an expectation value $\langle n(\mathbf{q}\neq0) \rangle, \langle p(\mathbf{q}\neq0) \rangle = 0$; on the cylinder, we check for peaks in the structure factors, shown in Fig.~\ref{fig:structure}.
The density-density correlations $S_{nn}(q) = \langle : n(\mathbf{q}) n(-\mathbf{q}) : \rangle_{\textrm{con}}$ do not show any delta-function like peaks, which suggests that $\langle n(\mathbf{q}\neq 0) \rangle = 0$ in the thermodynamic limit.
Indeed, $S_{nn}(q) $ changes little from $\delta=0$, which we know is a gapped liquid. 
The polarization - polarization correlation function $S_{pp}(q) = \langle : p(\mathbf{q}) p(-\mathbf{q}): \rangle_{\textrm{con}}$  is more interesting. 
 On the one-hand, $S_{pp}(q)$ also shows no delta-function like peaks.
However for  $\delta > 0$,  $S_{pp}(q)$ does appear to have non-analytic \emph{kinks}; in particular we will show that $S_{pp}(q) \sim |q|$ as $q \to 0$. 
 
 \subsection{Evidence for gapless polarization sector: finite entanglement scaling}
 
 	Admittedly, from the data of Fig.~\ref{fig:structure} the low-$q$ behavior of $S_{pp}$ looks rather smooth.
This is in fact an artifact of the finite bond dimension $\chi$ used in the DMRG simulations, which cuts off correlations at a ``finite entanglement'' correlation length $\xi(\chi)$,\cite{PollmannFES} and hence rounds out features in the structure factor at scale $1/\xi$.
However, conducting a ``finite entanglement scaling analysis'' in the bond dimension $\chi$ will allow us to demonstrate that $S_{pp}(q) \sim |q|$ as $\chi \to \infty$, as follows.

\begin{figure}
\includegraphics[width=1.\linewidth]{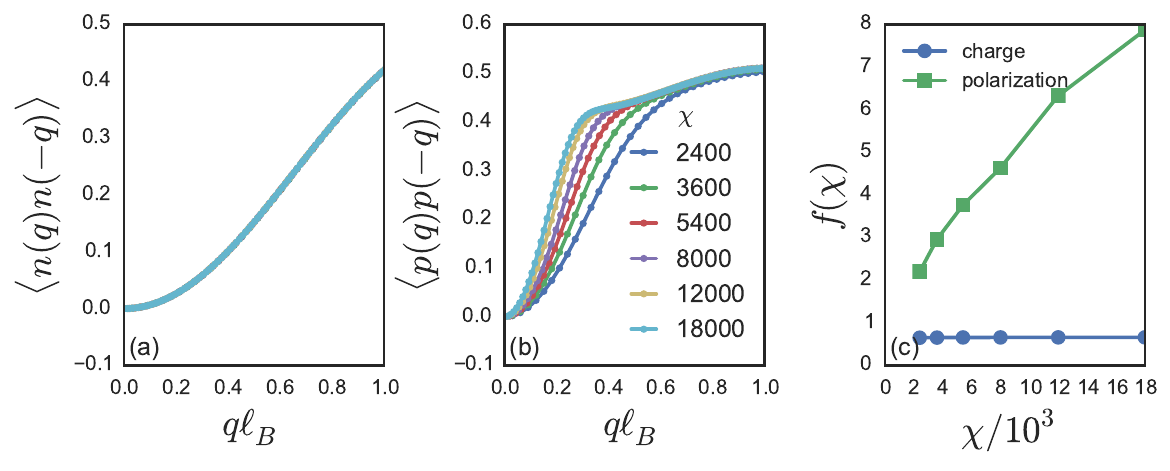}
\caption{Charge ({\bf left}) and polarization ({\bf middle}) correlators, similar to Fig. ~\ref{fig:structure}, but with fixed $\delta=1/12$ and varying DMRG bond dimension $\chi$ (note the values differ by a constant shift from Fig.~\ref{fig:structure}, as we have dropped the normal ordering).
The charge structure factor does not depend on $\chi$, while the $q\rightarrow 0$ behavior of the polarization structure factor gets sharper and sharper as $\chi$ is increased. This leads us to conclude that in the $\chi \to \infty$ limit the polarization correlations are singular at $q \to 0$ and the system  has a neutral gapless mode. {\bf Right}: We extract the $q=0$ curvature of the structure factor, $f(\chi) = \partial_q^2 S(q, \chi)|_{q=0}$. Following the scaling analysis of the main text, the large scaling of $f$ with $\chi$ indicates that the polarization mode is gapless, in contrast to the charge mode for which $f(\chi)$ is constant.
}
\label{fig:scaling}
\end{figure}

	In Fig.~\ref{fig:scaling} we show the evolution of the $q\sim 0$ behavior of the structure factors as the DMRG bond dimension $\chi$ is increased.
Indeed, while $S_{nn}$ is $\chi$ independent, the $S_{pp}$ correlations become sharper and sharper.
To analyze this scaling quantitatively, we assume the structure factor at bond dimension $\chi$,  $S(q; \xi(\chi))$, splits into an analytic part $S^{(a)}$ and a scaling part $S^{(s)}$. Since the scaling part of the structure factor should have scaling dimension 1 (i.e. $S^{(s)}(q, \xi) = \xi^{-1} S^{(s)}(q \xi, 1)$), this motivates a scaling collapse of the form
\begin{align}
\partial_q^2 S(q; \xi) |_{q = 0} \equiv f(\xi) = s^{(a)} + \xi s^{(s)}
\end{align}
If the polarization fluctuations are critical, then $s^{(s)} \neq 0$.
The result is shown in Fig.~\ref{fig:scaling}, and confirms that the charge correlations are  analytic at $q \to 0$, while the polarization sector is gapless,  $S_{pp}(q) \propto |q|$.

\subsection{Absence of ODLRO in the bosonic exciton correlations}

 \begin{figure}
\includegraphics[width=0.95\linewidth]{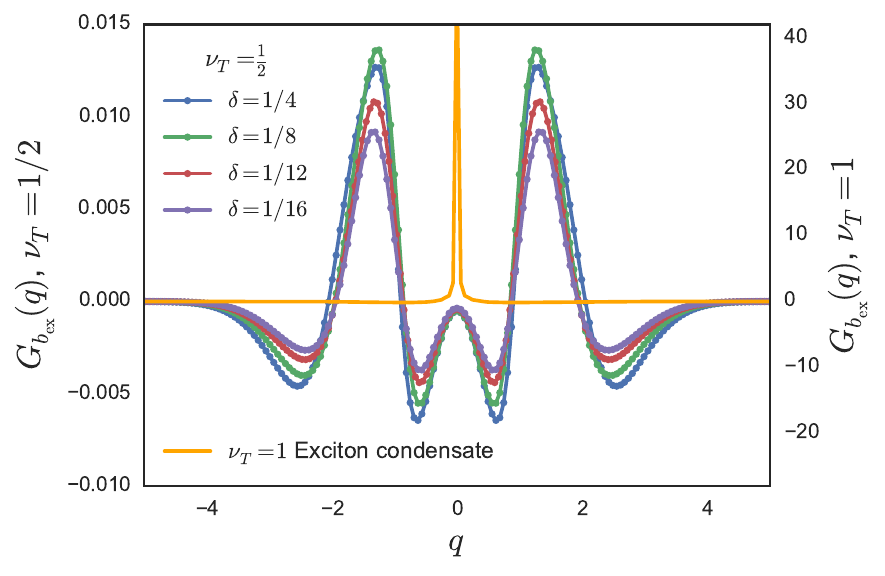}
\caption{Momentum-space bosonic exciton density $G_{\bex}(q)$. If a bosonic exciton condenses at momentum $q_\ast$, then $G_{\bex}(q)$ should show a peak at $q = q_\ast$ arising from algebraic ODLRO along the infinite cylinder.
For comparison,  the ``$\nu_T = 1$ exciton condensate'' data was obtained for a bilayer of two $N=0$ LLs with layer separation $d = 0.5 \ell_B$ and total filling $\nu_T=1$, where a $q_\ast = 0$ exciton condensate is known to occur ($L = 16, \chi = 5400$). The remaining curves are for the subject of this work: a bilayer of $N=1$ and $N=0$ LLs at filling $\nu_1 = \frac{1}{2} - \delta, \nu_0 = \delta$.
The peak at $q=0$ is completely absent. Note that the scale of the $y$-axis differs by three orders of magnitude for the two cases (annotated to the left and right).}
 \label{fig:exciton}
\end{figure}

	The  above results do not distinguish between an exciton metal and an exciton condensate, so to distinguish between the two  we measure the ``bosonic exciton correlator:''
\begin{equation}
G_{\bex}(q) \equiv \langle  : b^\dagger(\mathbf{q}) b(\mathbf{q}) : \rangle, \;\;\;\;\; b^\dagger(x) = \psi_0^\dagger(x) \psi_1(x).
\label{eq:gbex}
\end{equation}
In an exciton condensate, one expects a peak at the momentum $q=q_\ast$ of the condensate, while we do not expect a peak for the exciton metal.
In Fig.~\ref{fig:exciton} we plot $G_{\bex}(q)$. For contrast, we also consider a bilayer of two $N=0$ LLs at total filling $\nu_T=1$, which is known to exhibit an exciton condensate phase\cite{EisensteinReview}. At $\nu_T=1$,  $G_{\bex}(q)$ shows a singular peak at $q=0$, as expected. In this case, the system exhibits a linearly dispersing Goldstone mode, as has been shown in exact diagonalization simulations~\cite{Moon, PapicMeron}. On the other hand, at $\nu_T=\frac{1}{2}$, $G_{\bex}(q)$ is three orders of magnitude smaller and shows no such peak, which is strong evidence that the intermediate $\delta$ phase  is {\it not} an exciton condensate. 
Note that we cannot explicitly compute the analogous two-point function of the $\fex$ because it is a non-local excitation.

\section{Conclusion}\label{sec:conc}

We have given a microscopic picture for why the electric-field driven crossing of a $N=1$ and $N=0$ LL in bilayer graphene at $\nu_T = \frac{1}{2}$ filling should stabilize a new phase of matter, the topological exciton metal. Complementary exact diagonalization and DMRG calculations support the existence of this phase for a realistic model of  BLG.
	
Circumstantial  evidence for this phase - namely, the surprising coexistence of a quantized $\sigma_H = \frac{1}{2} \frac{e^2}{h}$ charge gap with a finite layer polarization at a crossing of $N=0$ and $N=1$ LLs - has already been obtained in experiment.\cite{Zibrov} However, these thermodynamic measurements were not sensitive to the differences between an exciton metal, exciton condensate or perhaps even phase separation.
Fortunately, the exciton metal would have dramatic transport signatures, such as metallic counterflow transport in a charge insulator.
Counterflow transport has already been used to detect the bosonic exciton condensate at $\nu_T=1$ in BLG.\cite{li2017excitonic} However, these results relied on a \emph{bilayer} of BLG - e.g., two sheets of BLG separated by a very thin ($d \sim 2.5$nm) boron-nitride spacer, with indirect excitons forming across the spacer. 
The bilayer of BLG is required because there is no obvious way to separately contact the two (atomically close) layers within a single sheet of BLG.
Luckily our scenario should also be realizable in the bilayer of BLG (or a bilayer of monolayer and bilayer graphene).
By using top and bottom gate electrodes one can engineer a crossing between a $N=0$ and a $N=1$ LL such that each is isolated in a \emph{different} BLG. For a thin boron nitride spacer, $ d \ll \ell_B$, so the fact that the two LLs are separated by a spacer, rather than within the same BLG, should not modify our analysis. We hope our results give a compelling reason to pursue this direction.

\acknowledgments
We are indebted to conversations with M. Barkeshli, R. Mong, C. Nayak,  A. Young and J. Zhang. The DMRG calculations were performed  on computational resources supported by the Princeton Institute for Computational Science and Engineering using iDMRG code developed with Roger Mong and the TenPy collaboration.
S.G and E.R. were supported by Department of Energy BES Grant DE-SC0002140. Z.P. acknowledges support by EPSRC grant EP/P009409/1. Statement of compliance with EPSRC policy framework on research data: This publication is theoretical work that does not require supporting research data.

\bibliography{case}

\appendix

\section{Additional DMRG data}

In this appendix we provide some additional details and DMRG data which support the conclusions in Section~\ref{sec:dmrg}. 

\subsection{Form factors used in the iDMRG simulations}

In bilayer graphene the form factor of the $N=1$ LL (and hence the effective interaction) takes the  general form \cite{papic2011tunable}
\begin{align}
F_{11}(q) = e^{-\frac{1}{2} q^2}( \cos^2(\Theta)(1 - q^2/2) + \sin^2(\Theta) )
\end{align}
where $\Theta$ depends on the magnetic field. $\Theta = 0$ corresponds to a conventional $N=1$ LL, while $\Theta = \pi / 2$ corresponds to the interactions of a conventional $N=0$ LL, which are sharper.  For  the fields $B = 0 - 15$T relevant to most experiments, $\Theta \sim 0 - 0.35$ \cite{hunt2017direct}.
Since a small $\Theta$ sharpens the interactions, for ``historical reasons'' we stabilized the Pfaffian  in our iDMRG simulations  by setting $\Theta = 0.1$, while in  ED we used the pseodopotential perturbation $0.05 V_1$.
While the interactions aren't identical, the difference only leads to small quantitative change in the energies.

\subsection{Fermi surface and central charge in iDMRG}

While ruling out several alternatives, unlike exact diagonalization the DMRG evidence does not directly provide a ``smoking-gun'' signature of an exciton FS.
For example, the exciton FS may compete with a two-component CFL (2CFL) formed when the CFs in both layers form Fermi surfaces with  volumes proportional to $\nu_{+1}$ and $\nu_{-0}$ respectively. 
Region II.2 of the BLG experiment,\cite{Zibrov} $0.18 < \delta < 0.5$, which is compressible and polarizable, may be such a 2CFL.
An obvious distinction between the exciton FS and 2CFL is the presence vs. absence of a charge gap, but our DMRG simulations only obtain the ground state.

Another sharp distinction is the volume of their Fermi surfaces,  $V_{\textrm{exFS}} = 2 \pi \delta \ell_B^{-2}$ vs. $ V_{\textrm{2CFL}}  = \pi (1 - 2 \delta) \ell_B^{-2} + \pi 2 \delta \ell_B^{-2}$.
One way to measure the Fermi volume is by analyzing  the non-analytic kinks in the $S_{pp}(q)$ structure factor of Fig.~\ref{fig:structure}(b), which occur at momenta $q^\ast$ corresponding to scattering across the FS.
In Ref.~\onlinecite{Geraedts} such information was used to map out the FS of the one component CFL.
An example of what we believe is such a singularity is the broad ``shoulder'' in $S_{pp}$, i.e. around $q^\ast \ell_B \sim 0.25$ for $\delta=1/12$. The $q^\ast$ of this shoulder increases with $\delta$, presumably as the radius of the $\fex$ FS increases.
However, because the massive entanglement in this system prevents us from reaching large finite-entanglement correlation lengths $\xi$, the feature remains broad, though signs of a kink can be further highlighted by taking  derivatives of the data. 
The difficulty is further exacerbated because the interplay of a small  $V_{\textrm{exFS}}$ and the quantizing effect of the cylinder circumference forces the FS to distort away from a circle.  We believe this places the backscattering wavevectors $q^\ast$ close together compared to the $\xi$-induced broadening, so we cannot make out the detailed structure of the FS.

\begin{figure}
\includegraphics[width=0.95\linewidth]{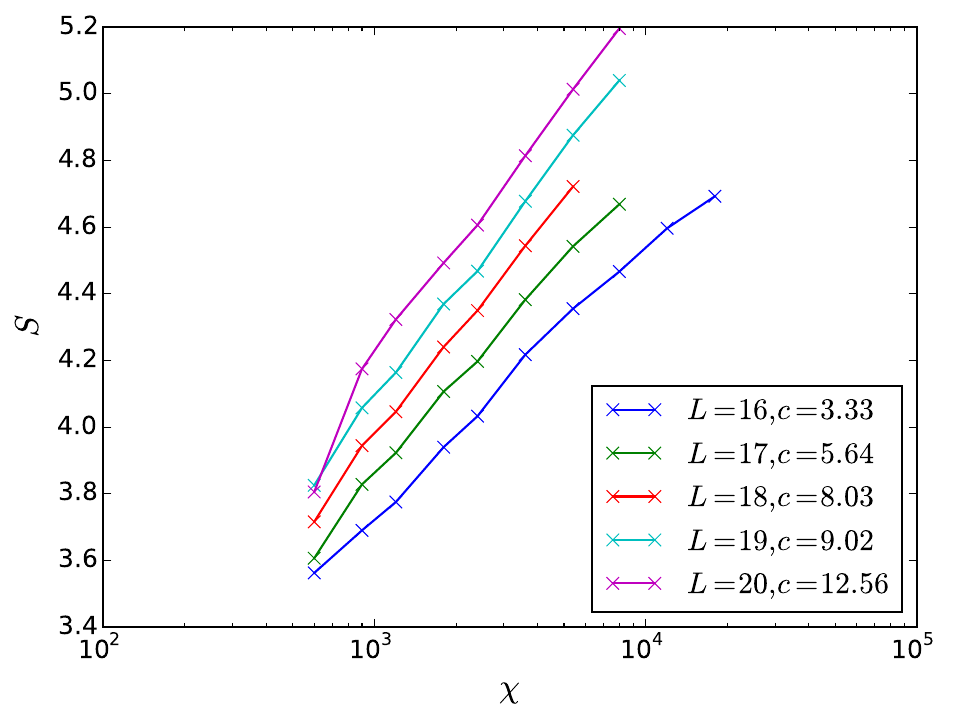}
\caption{Entanglement entropy, $S$, vs. the log of the bond dimension, $\chi$. We expect these quantities to obey Eq.~(\ref{Schi}). Our data shows a roughly linear relationship, as this equation would predict. Unfortunately even small errors in $S$ (such as those caused by finite-size effects) can significantly change the estimate of $c$ extracted with this method. The caption shows the values of $c$ extracted. We expect that regardless of what state we are realizing (i.e. exciton metal, exciton condensate or composite Fermi liquid) $c$ should stay constant for nearby values of $L$, occasionally increasing in steps of $2$. The fact that we do not see this behavior leads us to conclude that finite-size effects are altering $c$. This makes it impossible for us to use $c$ to determine what phase we are realizing. Data was taken with $\delta=1/12$, bond dimension up to $18000$.
}
\label{fig:central_charge}
\end{figure}

Alternatively, we can detect the FS by measuring the dependence of the central charge $c$ with cylinder circumference, $c \propto L$.\cite{Geraedts}  We  attempted to measure the central charge in our exciton system, extracted using the following finite-entanglement scaling (FES) formulae:
\begin{eqnarray}
&&S=\frac{c}{6}\log(\xi) \label{Sxi} \\
&& S=\frac{1}{\sqrt{\frac{12}{c}}+1} \log(\chi). \label{Schi}
\end{eqnarray}
Of the two formulae  Eq.~(\ref{Sxi}) is usually more numerically stable, but unfortunately at the bond dimensions available we cannot measure $\xi$ (the correlation length) accurately enough to use it.
This leaves us with Eq.~(\ref{Schi}), and we plot $S$ vs. $\log(\chi)$ in Fig.~\ref{fig:central_charge}. The problem with doing this is that due to the form of Eq.~(\ref{Schi}) small changes in the slope lead to large changes in $c$, especially when $c$ is reasonably large. The exciton condensate should have $c=1$, while the exciton metal phase can have $c\approx 2-5$, depending on $L$ (and $c$ should remain constant for several $L$ before jumping by $2$ approximately every $\Delta L\approx 2\pi/\delta$\cite{Geraedts}. 
Our data does not appear to do either, leading us to believe $\chi$ is too small to estimate the central charge: the $S(\chi)$ curves are noisy and not straight (compare with Ref.~\onlinecite{Geraedts}), which  suggests we are not in the finite-entanglement scaling regime.
Since the Pfaffian itself requires $\chi \approx 6000$, and we can only triple this, this  isn't so surprising.

\end{document}